\documentclass[aps,prb,twocolumn,superscriptaddress,floatfix]{revtex4-1}

\usepackage{graphicx,graphics}
\usepackage{dcolumn}
\usepackage{amsmath,amssymb,amsfonts}
\usepackage{latexsym,verbatim}
\usepackage{bm}
\usepackage{color}
\usepackage{ulem}
\usepackage[percent]{overpic}
\usepackage[breaklinks=true,colorlinks,citecolor=blue,linkcolor=blue,urlcolor=blue]{hyperref}

\begin{document}
\title{Scaling approach to tight-binding transport in realistic graphene devices: the case of transverse magnetic focusing}
\author{M. Beconcini}
\email{michael.beconcini@sns.it}
\affiliation{NEST, Scuola Normale Superiore, I-56126 Pisa,~Italy}
\author{S. Valentini}
\affiliation{NEST, Istituto Nanoscienze-CNR and Scuola Normale Superiore, I-56126 Pisa,~Italy}
\author{R. Krishna Kumar}
\affiliation{Physics Department, Lancaster University, Lancaster LA14YB,~United Kingdom}
\affiliation{School of Physics \& Astronomy, University of Manchester, Oxford Road, Manchester M13 9PL,~United Kingdom}
\author{G.H. Auton}
\affiliation{National Graphene Institute, University of Manchester, Manchester M13 9PL,~United Kingdom}
\author{A.K. Geim}
\affiliation{School of Physics \& Astronomy, University of Manchester, Oxford Road, Manchester M13 9PL,~United Kingdom}
\author{L.A. Ponomarenko}
\affiliation{Physics Department, Lancaster University, Lancaster LA14YB,~United Kingdom}
\affiliation{School of Physics \& Astronomy, University of Manchester, Oxford Road, Manchester M13 9PL,~United Kingdom}
\author{M. Polini}
\affiliation{Istituto Italiano di Tecnologia, Graphene Labs, Via Morego 30, I-16163 Genova,~Italy}
\author{F. Taddei}
\affiliation{NEST, Istituto Nanoscienze-CNR and Scuola Normale Superiore, I-56126 Pisa,~Italy}
\begin{abstract}
Ultra-clean graphene sheets encapsulated between hexagonal boron nitride crystals host two-dimensional electron systems in which low-temperature transport is solely limited by the sample size. We revisit the theoretical problem of carrying out microscopic calculations of non-local ballistic transport in such micron-scale devices. By employing the Landauer-B\"uttiker scattering theory, we propose a novel scaling approach to tight-binding non-local transport in realistic graphene devices. We test our numerical method against experimental data on transverse magnetic focusing (TMF), a textbook example of non-local ballistic transport in the presence of a transverse magnetic field. This comparison enables a clear physical interpretation of all the observed features of the TMF signal, including its oscillating sign.
\end{abstract}
\maketitle

\section{Introduction}
\label{sect:intro}

The ability to fabricate ultra-clean graphene sheets by encapsulation in hexagonal boron nitride crystals~\cite{mayorov_nanolett_2011,wang_science_2013,taychatanapat_naturephys_2013,bandurin_science_2016} allows the investigation of ballistic transport in a large range of temperatures up to the hydrodynamic temperature scale~\cite{bandurin_science_2016,torre_prb_2015} $T_{\rm hydro}$. At $T\gtrsim T_{\rm hydro}$, the mean-free path for electron-electron collisions $\ell_{\rm ee}$ becomes shorter than the mean free path $\ell$ for momentum non-conserving scattering and inelastic electron-electron collisions need to be taken into account in any theoretical description of transport. 

\begin{figure}[h!]
\begin{overpic}[width=0.49\linewidth]{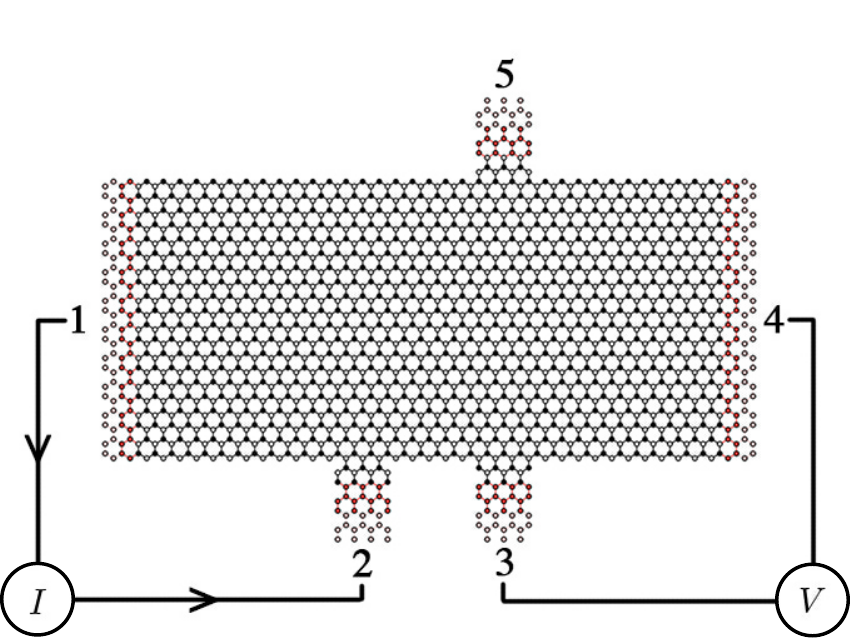}\put(2,62){(a)}\end{overpic}
\begin{overpic}[width=0.49\linewidth]{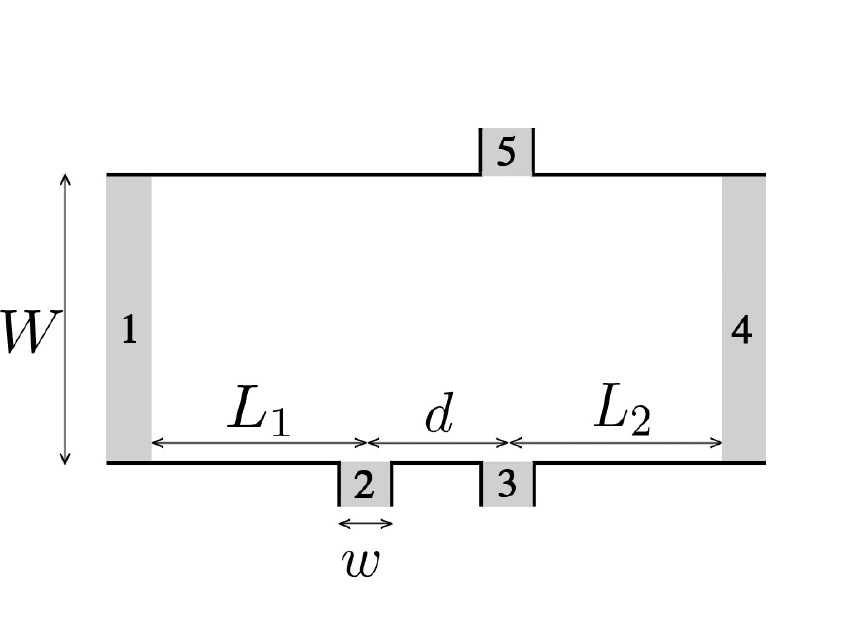}\put(2,62){(b)}\end{overpic}
\begin{overpic}[width=0.49\linewidth]{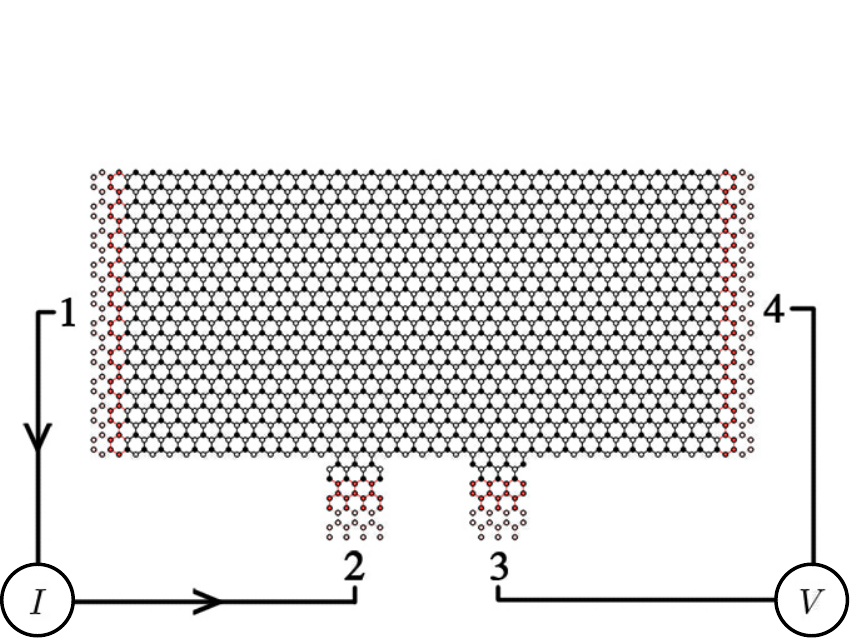}\put(2,62){(c)}\end{overpic}
\begin{overpic}[width=0.49\linewidth]{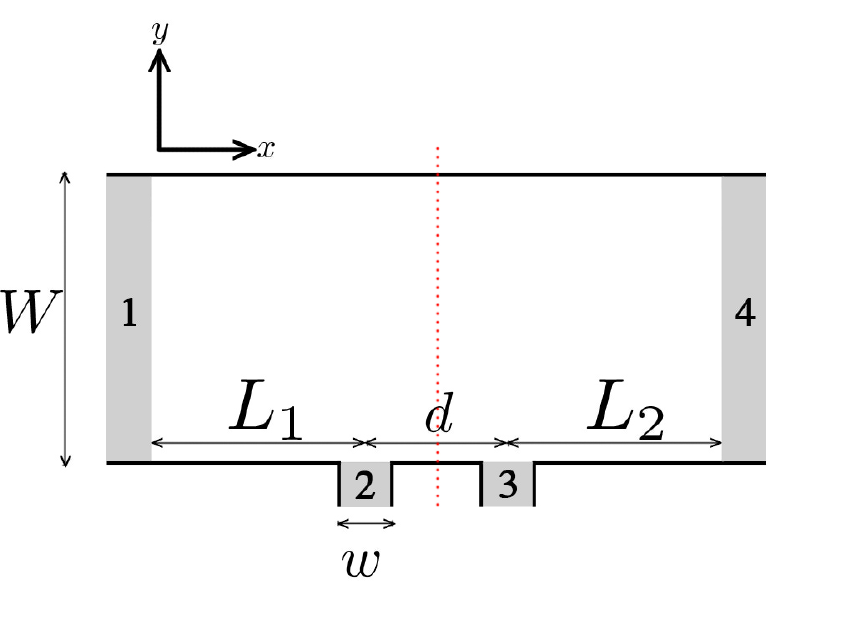}\put(2,62){(d)}\end{overpic}
\begin{overpic}[width=0.49\linewidth]{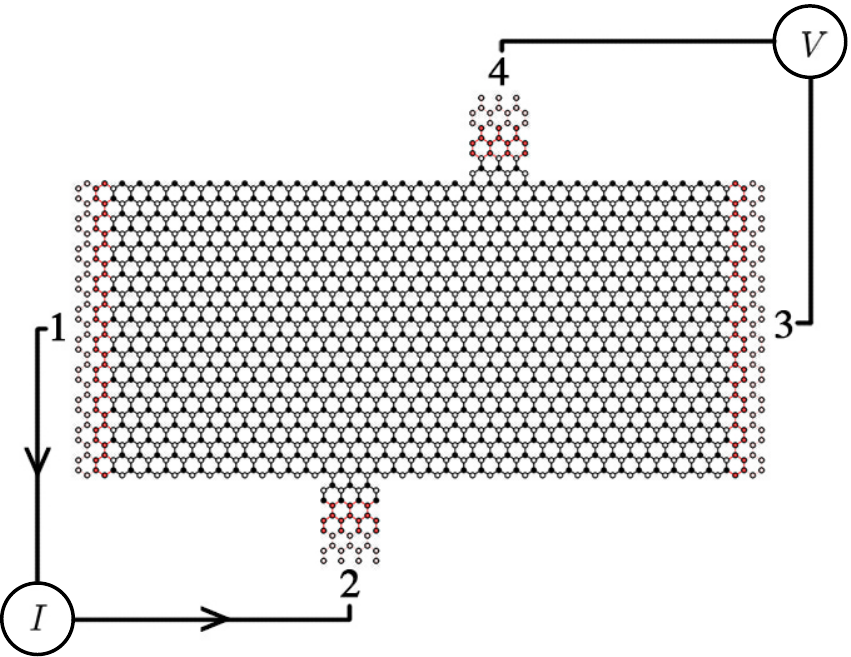}\put(2,62){(e)}\end{overpic}
\begin{overpic}[width=0.49\linewidth]{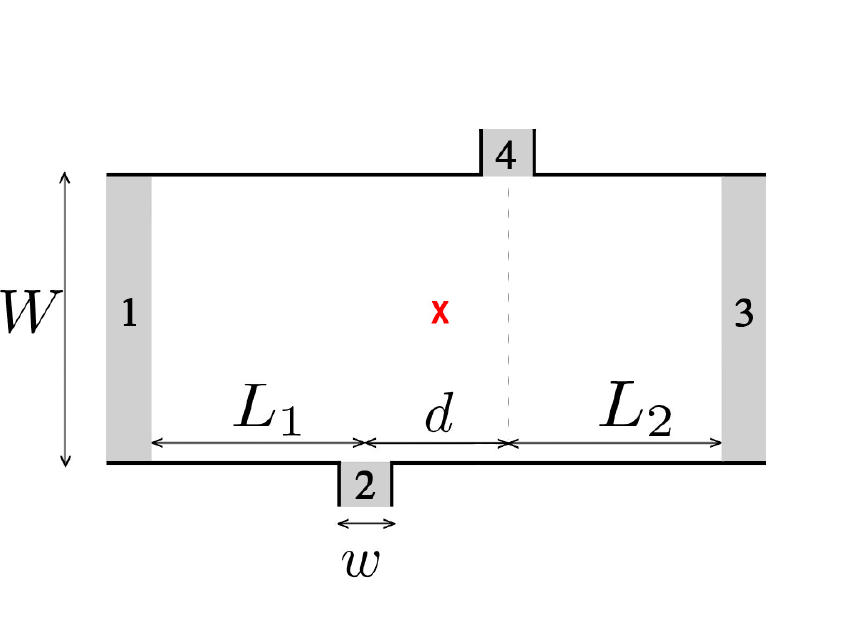}\put(2,62){(f)}\end{overpic}
\caption{(Color online) Pictorial representation of the five and four-terminal graphene Hall bar setups considered in this work. Leads are labeled by  numbers. Here, $W$ is the width of the horizontal zig-zag leads, $w$ is the width of the vertical armchair leads, $d$ is the center-to-center distance between them, $L_1$ is the distance between lead 1 and the center of lead 2, and $L_2$ is the distance between the center of lead 3 and lead 4, for panel (b) and (d), or the distance between the center of lead 4 and lead 3, for panel (f). The red dotted line (red cross) in panel (d) (panel (f)) indicates the axis (center) of symmetry of the setup. In this Article, all leads have been taken to be semi-infinite. \label{fig:one}}
\end{figure}

At temperatures $T\ll T_{\rm hydro}$, however, electrons in encapsulated graphene sheets propagate over distances of the order of several microns without experiencing elastic or inelastic scattering events. In this situation, transport properties can be determined by utilizing exact single-particle quantum approaches, combining e.g.~tight-binding Hamiltonians with Kubo formulas~\cite{yuan_prb_2010,roche_ssc_2012} or  Landauer-B\"uttiker scattering theory~\cite{kwant,liu2015}. 

Graphene Hall bars fabricated by van der Waals assembly techniques and used in quantum transport experiments have characteristic linear dimensions of tens of microns, rendering brute-force numerical calculations time consuming or, simply, unfeasible. In Ref.~\onlinecite{liu2015} a convenient scaling scheme for two-terminal numerical transport simulations within Landauer-B\"uttiker scattering theory has been proposed. In such a scheme, the tight-binding parameters for real graphene~\cite{neto_rmp_2009}, namely the hopping energy $t_0$ and the lattice spacing $a_{0}$, are replaced with rescaled ones, $\tilde{t}_{0}$ and $\tilde{a}_{0}$, such that the bulk band structure $E(k)$ remains invariant, i.e.~ $E(k)=(3/2)t_{0}a_0k=(3/2)\tilde{t}_{0} \tilde{a}_{0} k$, with $k$ the magnitude of the momentum. This yields the scaling condition $\tilde{a}_{0}=a_{0}s_{\rm f}$ and $\tilde{t}_{0}=t_{0}/s_{\rm f}$, which applies only when the massless Dirac (linear) approximation is valid, where $s_{\rm f}$ is the scaling factor. Restrictions for the validity of the scaling procedure, in terms of a maximum scaling factor, are derived on the basis of the {\it bulk} band structure. As an example, this scaling procedure has been used~\cite{liu2015} to simulate the two-terminal conductance, measured on a large crystal, using a scaling factor up to $100$.

In this Article we develop a scaling procedure suitable for graphene micron-sized ribbons, which is valid also in the presence of many electrodes (see Fig.~\ref{fig:one}) and therefore useful to describe {\it non-local} ballistic transport experiments. This procedure is based on the exact band structure of graphene ribbons (rather than on the bulk massless Dirac fermion band structure), and uses the Fermi energy as key scaling parameter. In brief, a geometrical downward scaling of the size of the structure, from the realistic laboratory scale to the computationally feasible scale, is accompanied by a upward scaling of the Fermi energy in such a way that the number of electronic modes responsible for transport is left unchanged. 

As an application of the proposed scaling procedure, we study in detail the case of transverse magnetic focusing (TMF), which has been extensively explored in the past, in metals~\cite{tsoi_jetp_1974} and in ultra-clean semiconductor heterostructures~\cite{vanhouten_epl_1988,vanhouten_prb_1989,heremans_prb_1995,heindrichs1998,rokhinson_prl_2004,tsoi_rmp_1999} fabricated by molecular beam epitaxy. Here, we focus on TMF in single-layer graphene~\cite{taychatanapat_naturephys_2013,bhandari_nanolett_2016}, comparing our quantum mechanical numerical calculations with experimental results in ultra-clean encapsulated monolayer samples.

Our paper is organized as following. In Sect.~\ref{sect:TB} we present our scaling approach. In Sect.~\ref{sect:numerical_examples} we summarize our main numerical results on TMF in single-layer graphene, while Sect.~\ref{analy} is devoted to a detailed analysis of the numerical results. In particular, Sect.~\ref{analy} includes a study of the dependence of TMF on the carrier density, temperature, presence of non-ideal edges, as well as a comparison with experimental data. A brief summary and our main conclusions are reported in Sect.~\ref{sect:conc}.

\section{Theoretical framework and scaling procedure}
\label{sect:TB}
The systems under investigation are multi-terminal graphene Hall bars similar to the ones sketched in Fig.~\ref{fig:one}. A rectangular graphene zig-zag strip, of width $W$, is attached either to $5$~[Figs.~\ref{fig:one}(a)-(b)] or $4$~[Figs.~\ref{fig:one}(c)-(d) and~\ref{fig:one}(e)-(f)] electrodes and exposed to a perpendicular magnetic field ${\bm B}$. The horizontal leads [labeled 1 and 4 in Figs.~\ref{fig:one}(a)-(d), and 1 and 3 in Figs.~\ref{fig:one}(e)-(f)] have the same width $W$ of the ribbon, while the vertical ones have width $w \ll W$. In our calculations below, all leads have been taken to be semi-infinite. Moreover, the vertical terminals are separated by a center-to-center distance $d$, while the distance between the left (right) horizontal electrode and the leftmost (rightmost) vertical electrode is $L_1$ ($L_2$)  [see Figs.~\ref{fig:one}(b), (d) and~(f)].
The total length of the ribbon is therefore $L=L_1+d+L_2$.

In all setups a non-local resistance $R_{21,34}$ is measured by applying a current bias between lead 1 and 2 and measuring the voltage that develops between lead 3 and 4. We therefore define
\begin{equation}\label{eq:non-local}
R_{21,34}=\frac{V_3-V_4}{I_2}~,
\end{equation}
where $I_i$ is the current flowing in lead $i$ and $V_i$ is the voltage relative to lead $i$.
As we mentioned in the Introduction, in high-quality encapsulated graphene we can safely assume that low-temperature transport is coherent and neglect inelastic scattering sources. For the sake of simplicity, we also neglect elastic scattering sources: our work does not therefore deal with carrier density inhomogeneities near the charge neutrality point.

The single-particle tight-binding Hamiltonian reads
\begin{equation}\label{tbHam}
{\cal H} = \varepsilon_{\rm F} \sum_i  c^{\dagger}_i c_i -t_0 \sum_{\langle i,j \rangle} c^{\dagger}_i c_j~,
\end{equation}
where $\varepsilon_{\rm F}$ is the Fermi energy, measured with respect to the Dirac point ($\varepsilon_{\rm F}=0$), and $t_0\simeq 2.8~{\rm eV}$ is the nearest neighbor hopping energy (the symbol ${\langle i,j \rangle}$ denotes, as usual, nearest-neighbor sites $i$ and $j$). We remind the reader that electron-electron interactions, which are not included in our model Hamiltonian (\ref{tbHam}), enhance the value of the Fermi velocity~\cite{kotov_rmp_2012} $v_{\rm F}$ with respect to the bare non-interacting tight-binding value  $v_{{\rm F}, 0} = (3/2)t_{0}a_{0} \simeq 0.9 \times 10^{6}~{\rm m}/{\rm s}$.

The non-local resistance $R_{21,34}$ can be calculated starting from the linear-response current-voltage relation obtained within the Landauer-B\"uttiker scattering approach and given by\cite{buttiker1986,buttiker1988}
\begin{equation}\label{eq:currents}
I_i = \frac{2e^2}{h} \left[ (N_i-T_{ii})V_i - \sum_{j \ne i} T_{ij} V_j \right]~,
\end{equation}
at zero temperature. $R_{21,34}$ is obtained by imposing that $I_1=I_2$, $I_3=I_4=I_5=0$, and solving Eq.~(\ref{eq:currents}) for $V_3$ and $V_4$.
In Eq.~(\ref{eq:currents}) $T_{ij}$ is the transmission coefficient at the Fermi energy for electrons injected from lead $j$ to be transmitted into lead $i$, satisfying the identity
\begin{equation}\label{eq:sumrules}
N_i = \sum_j T_{ij} = \sum_j T_{ji}~,
\end{equation}
$N_i$ being the number of open channels in lead $i$.

The transmission coefficients $T_{ij}$ will be numerically calculated using KWANT\cite{kwant}, a toolkit which implements a wave-function matching technique.
We assume that no magnetic field is present in the leads.

\begin{figure}[t]
\begin{overpic}[width=1.0\linewidth]{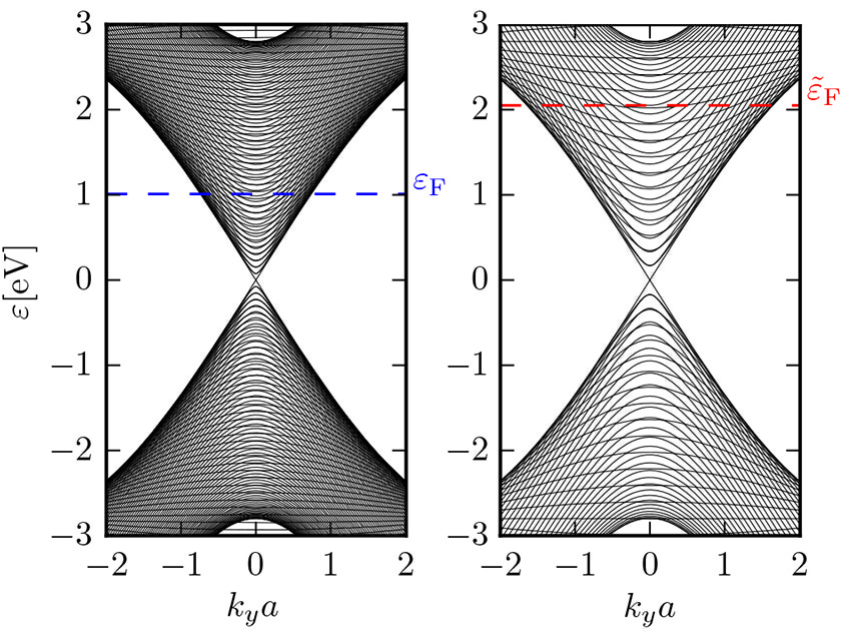}\put(0,70){(a)}\end{overpic}
\begin{overpic}[width=1.0\linewidth]{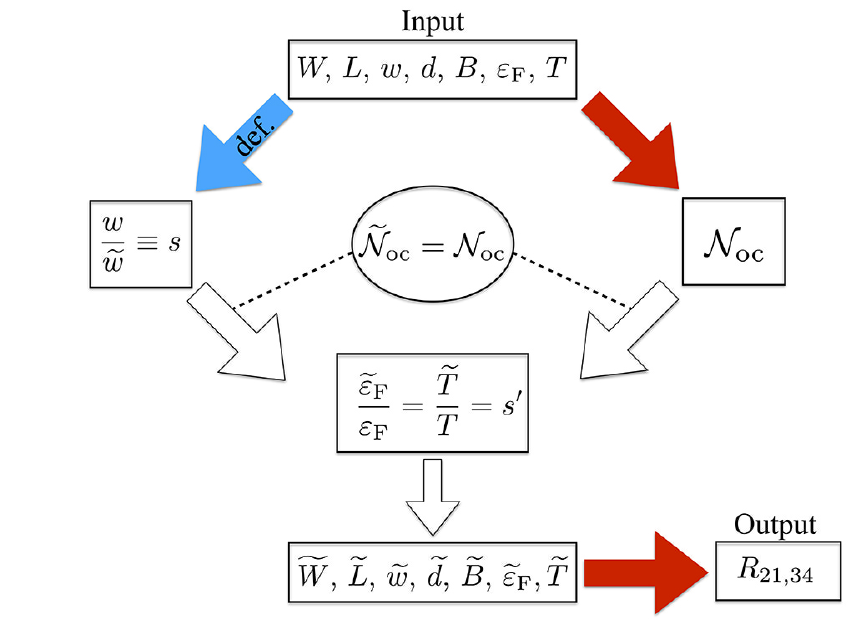}\put(0,70){(b)}\end{overpic}
\caption{\label{fig:two}
(Color online) (a) Two examples of band structures of {\it armchair} leads. Energies are measured in ${\rm eV}$, while $k_y$ is measured in units of $a = a_{0} \sqrt{3}$. On the left, $w = 24.4~{\rm nm}$ and $\varepsilon_{\rm F}=1.01~{\rm eV}$; on the right, $\tilde{w} = 10.8~{\rm nm}$ and $\tilde{\varepsilon}_{\rm F}=2.05~{\rm eV}$. (b) Schematic representation of the scaling procedure. In the input box we have the parameters characterizing the real sample: $W$, $L$, $w$, $d$, $B$, $\varepsilon_{\rm F}$, and $T$ (where $T$ is temperature). ${\cal N}_{\rm oc}$ is the number of open channels in a reference lead of the real sample. Quantities denoted by a tilde refer to the rescaled system. 
The parameters $s$ and $s'$ are the geometric and energy scaling factors, respectively. The rescaled parameters are used to calculate non-local resistances. In this work we focus on the quantity $R_{21,34}$ defined in Eq.~(\ref{eq:non-local}).}
\end{figure}

Since the computation time scales roughly with the third power of the linear size of the system\cite{kwant}, a one-to-one simulation of a large-size sample, of the order of a few micrometers, is prohibitively time consuming.
For this reason, the development of scaling procedures, which allow to calculate accurately the transmission coefficients on a much smaller sized system, is of great interest.

Here we develop a procedure which is based on the observation that the band structure of a graphene nanoribbon varies little if its width is decreased by a scaling factor $s$ and, at the same time, the Fermi energy is increased by a suitable and, in principle, different factor $s'$.
This is graphically exemplified in Fig.~\ref{fig:two}(a) where the band structures~\cite{akhmerov_prb_2008,katsnelson_book} of two armchair nanoribbons of different width (scaled by a factor $s \simeq 2.26$) are plotted side by side. The two plots resemble each other as long as the Fermi energy of the narrower nanoribbon is increased by a suitable factor $s'$. The notion of ``suitability'' will be clarified below. Note that this works as long as the Fermi energy in the right panel of Fig.~\ref{fig:two}(a) satisfies the inequality 
$\tilde{\varepsilon}_{\rm F} < t_{0}$. Given a certain Fermi energy $\varepsilon_{\rm F}$ relative to the actual sample, this sets a limitation on the maximum scaling factor $s$ applicable.

The scaling procedure is schematized in Fig.~\ref{fig:two}(b). The ``input'' block contains all the parameters characterizing the actual sample one is interested in simulating. The scaling algorithm proceeds as following.  i) One starts by choosing the size $\tilde{w}$ of the vertical leads of the rescaled system used in the calculations; ii) one then defines the {\it geometric} scaling factor $s$ (blue arrow), i.e.~the original width $w$ of the vertical leads in units of $\tilde{w}$; the procedure of geometric scaling, although applied to all the sample, is based on the vertical leads in Fig.~\ref{fig:one} since those are the narrowest ones; iii) by knowing $w$ and $\varepsilon_{\rm F}$, one proceeds by calculating the number of open channels in the actual sample (red arrow), which we denote by ${\cal N}_{\rm oc}$; iv) one then determines the {\it energy} scaling factor $s'$ by imposing that the number of open channels $\tilde{\cal N}_{\rm oc}$ in the rescaled system equals ${\cal N}_{\rm oc}$ (white arrows);  
v) the rescaled parameters (denoted by a tilde) are used to determine the transmission coefficients of the rescaled system and therefore the non-local resistance $R_{21,34}$. 

Note that the rescaled magnetic field $\tilde{B}$ is given by $\tilde{B} = s^2B$ to make sure that the flux is invariant under geometric scaling. 

\section{Numerical results for TMF}
\label{sect:numerical_examples}
In this Section we present numerical results based on the scaling procedure described above. We have decided to focus our attention on TMF.

We consider the 5-terminal setup in Fig.~\ref{fig:one}(a) and (b). For an armchair lead of width $w=0.37~{\rm \mu m}$ and a Fermi energy $\varepsilon_{\rm F} = 66.86~{\rm meV}$, we find a number ${\cal N}_{\rm oc}$ of open channels given by ${\cal N}_{\rm oc} =27$. This should be compared with the approximate formula
\begin{equation}\label{eq:Noc}
{\cal N}_{\rm oc} \simeq {\rm int} \left[ (2w+a_{0})\frac{2\varepsilon_{\rm F}}{hv_{\rm F}} \right]~,
\end{equation}
which was derived by using the Dirac equation with appropriate boundary conditions~\cite{brey2006}. For the parameters reported above, Eq.~(\ref{eq:Noc}) yields ${\cal N}_{\rm oc} = 26$. The difference is due to residual finite-size effects that are not captured by Eq.~(\ref{eq:Noc}).

\begin{figure}[h!]
\begin{overpic}[width=0.8\linewidth]{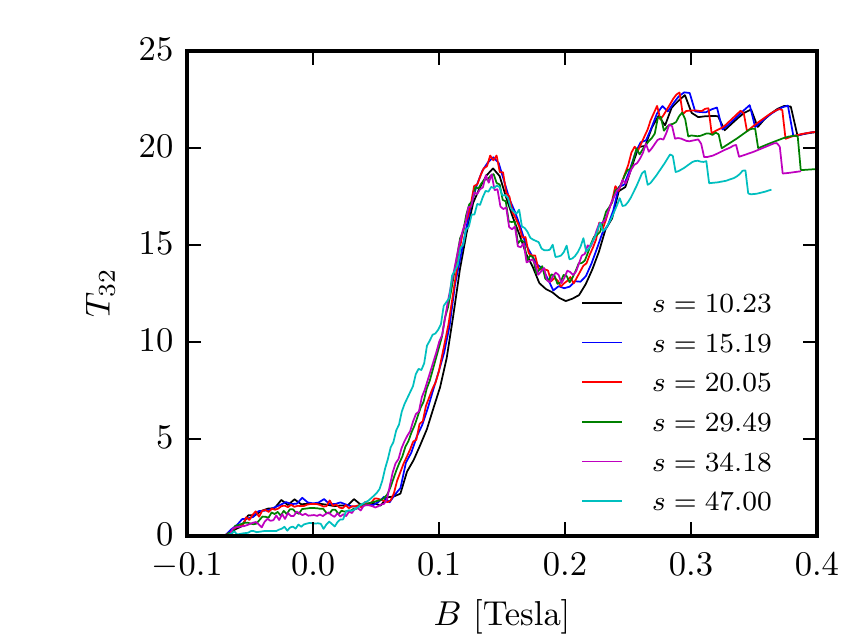}\put(2,70){(a)}\end{overpic}\\
\begin{overpic}[width=0.8\linewidth]{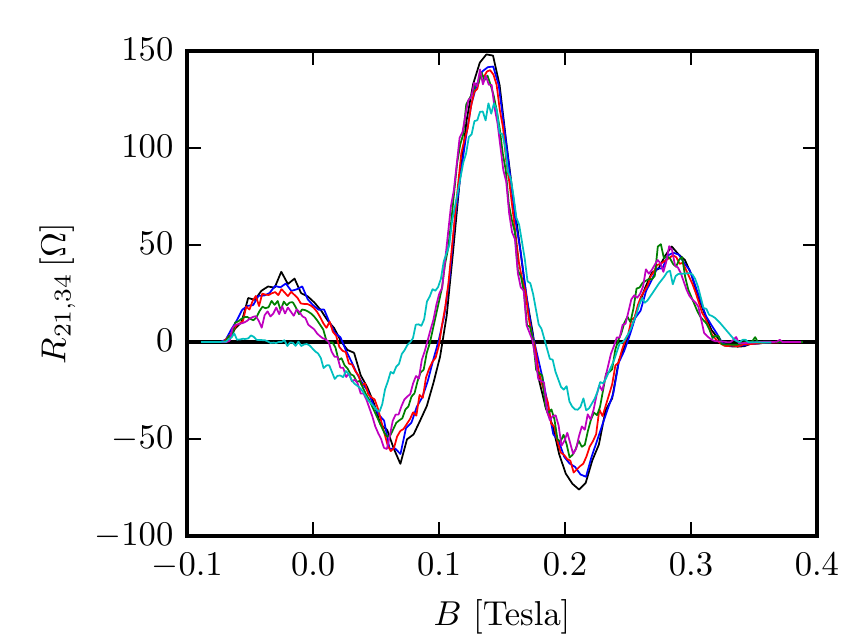}\put(2,70){(b)}\end{overpic}
\begin{overpic}[width=0.8\linewidth]{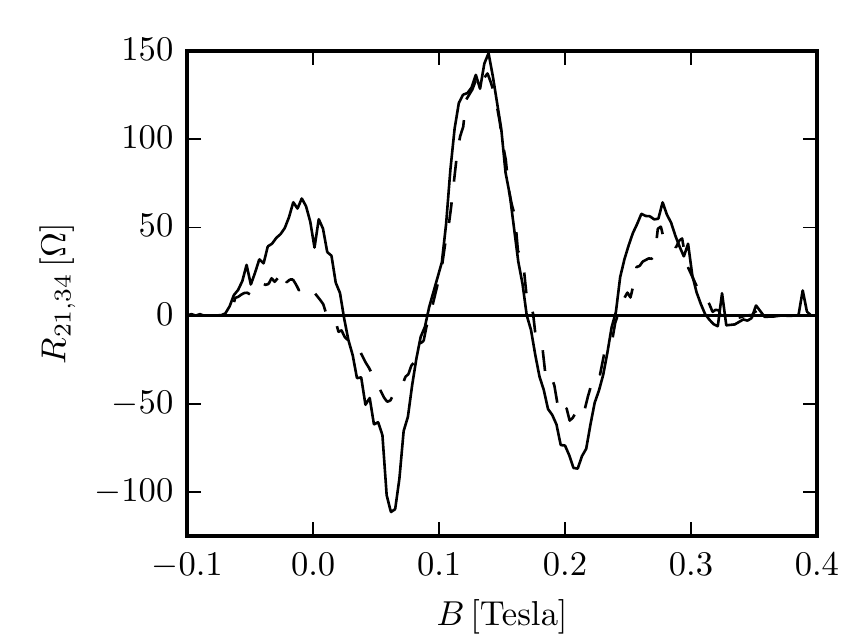}\put(2,70){(c)}\end{overpic}
\caption{\label{fig:three}
(Color online) (a) and (b) Numerical results for the transmission $T_{32}$---panel (a)---and the non-local resistance $R_{21,34}$---panel (b)---are plotted versus the applied magnetic field $B$ (in Tesla). Different curves refer to different values of the geometric scaling factor $s$ in the 5-terminal setup sketched in Figs.~\ref{fig:one}(a) and (b). The scaling procedure works well for $s\lesssim 34$. Numerical data presented in this figure were obtained for the following choice of parameters: $W=2~{\rm \mu m}$, $L_1=L_2=1.5~{\rm \mu m}$, $w=0.37~{\rm \mu m}$, $d=1~{\rm \mu m}$, $\varepsilon_{\rm F} = 66.86~{\rm meV}$ and ${\cal N}_{\rm oc} = 27$.
(c) Non-local resistance $R_{21,34}$ versus $B$ for an armchair (solid line, scaling factor $s=29.50$, number of open channels in lead $2$: ${\cal N}_{\rm oc} = 25$) and a zig-zag  (dashed line, scaling factor $s=29.49$, number of open channels in lead $2$: ${\cal N}_{\rm oc} =27$) ribbon. The energy scaling factor is $s' = 21.06$ in both cases.}
\end{figure}

To prove the effectiveness of the scaling procedure, we have compared the transmission $T_{32}$ and the non-local resistance $R_{21,34}$ at zero temperature for increasing values of the scaling parameter $s$ in Fig.~\ref{fig:three}.
The transmission $T_{32}$, plotted in Fig.~\ref{fig:three}(a) as a function of the magnetic field $B$, relative to electrons injected from lead 2 and arriving in lead 3, is the most relevant since it determines the main peak in the non-local resistance (see Sect.~\ref{analy}).

First, we notice that all curves in Fig.~\ref{fig:three}(a) do not show important quantitative differences up to $s\lesssim 34$, at least for fields as large as $0.3~{\rm Tesla}$. Similarly, Fig.~\ref{fig:three}(b) shows that the non-local resistance $R_{21,34}$ is only weakly sensitive to the scaling factor $s$. We also checked that the scaling procedure works well when the graphene Hall bar is an armchair ribbon, so that the vertical leads have zig-zag edges. In Fig.~\ref{fig:three}(c) we compare the non-local resistances $R_{21,34}$ of armchair (solid line) and zig-zag (dashed line) ribbons using approximatively the same geometric scaling factor $s \simeq 29.5$. Note that the two curves have the same behavior, the main focusing peak being virtually identical.

\begin{figure}[t]
\includegraphics[width=1.0\linewidth]{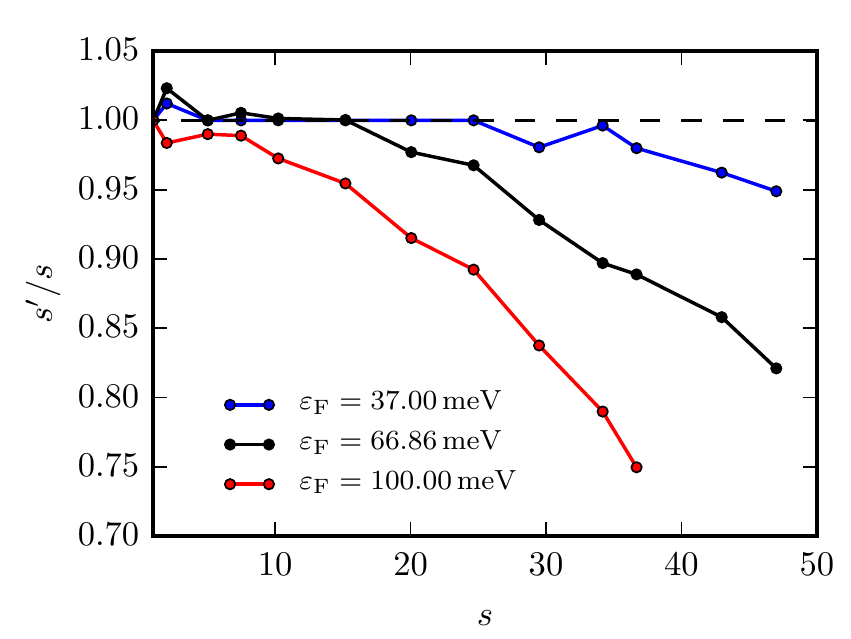}
\caption{\label{fig:four}
(Color online) Trend of the ratio $s'/s$ versus the geometric scale factor $s$ for three different values of the unscaled Fermi energy $\varepsilon_{\rm F}$. For the largest Fermi energy ($\varepsilon_{\rm F}=100$ meV), $s=36.68$ is the maximum value for which the scaling procedure can be applied (i.e. the  number of channels in the vertical leads can be kept fixed).}
\end{figure}

In Fig.~(\ref{fig:four}) we also present how the energy scaling factor $s'$ depends on $s$ for different values of the Fermi energy $\varepsilon_\text{F}$.
As expected, the plot shows that $s'$ tends to deviate from $s$ for large values of $s$ and more rapidly for large values of $\varepsilon_\text{F}$. We note that the functional dependence of $s'$ on $s$ is crucial. It makes sure that the position of the focusing peaks in the non-local resistance $R_{21,34}$---see Section~\ref{analy}---is insensitive to the geometric scaling factor $s$. Imposing that the rescaled system and the original one have the same number of open channels in the injection lead---$\tilde{\cal N}_{\rm oc} = {\cal N}_{\rm oc}$ through the parameter $s'$---guarantees that the rescaled-system band structure faithfully reflects  the original one.

\section{Detailed analysis of the numerical results}
\label{analy}

In this Section we analyze the origin of the different peaks exhibited by the non-local resistance as a function of $B$ and the origin of the sign of $R_{21,34}$ at {\it zero} magnetic field.

For the sake of simplicity, we consider the 4-terminal setup in Figs.~\ref{fig:one}(c) and~(d), where the zero-temperature non-local resistance is given by the following analytical expression~\cite{buttiker1986,buttiker1988}:
\begin{equation}\label{r2134}
R_{21,34}  = \frac{h}{2e^2}\frac{T_{32}T_{41}-T_{42}T_{31}}{D}~,
\end{equation}
where
\begin{equation}
D \equiv (\alpha_{11}\alpha_{22}-\alpha_{12}\alpha_{21})S~,
\label{di}
\end{equation}
\begin{equation}
S \equiv T_{13}+T_{14}+T_{23}+T_{24} = T_{31}+T_{41}+T_{32}+T_{42}~,
\end{equation}
\begin{align}
	\alpha_{11} &= \frac{2e^2}{h}\left[ T_1-\frac{(T_{13}+T_{14})(T_{41}+T_{31})}{S} \right]\\
	\alpha_{12} &= -\frac{2e^2}{h} \frac{T_{14}T_{23}-T_{13}T_{24}}{S} \\
	\alpha_{21} &= -\frac{2e^2}{h} \frac{T_{32}T_{41}-T_{42}T_{31}}{S} \\
	\alpha_{22} &= \frac{2e^2}{h}\left[ T_4-\frac{(T_{14}+T_{24})(T_{41}+T_{42})}{S}\right]~, \label{last}
\end{align}
and
\begin{equation}
T_{i}=\sum_{j\ne i} T_{ij}~.
\end{equation}

We start by discussing the behavior of the different transmission coefficients $T_{ij}$ as functions of the applied magnetic field in terms of a semiclassical picture and then see how they combine to give rise to the non-local resistance $R_{21,34}$ with the aid of Eq.~(\ref{r2134}).

Within a simple classical picture~\cite{heindrichs1998,milovanovic_jap_2014}, which will be corroborated below in Sect.~\ref{sect:classical}, electrons entering the Hall bar from a given electrode undergo a cyclotron motion with radius $r_{\rm c}=m^*v_{\rm F}/(eB)$ and specular reflections at the boundaries of the Hall bar.
In graphene, the cyclotron radius for weak magnetic fields can be written as
\begin{equation}\label{eq:cyclotronradius}
r_{\rm c} = \frac{\varepsilon_{\rm F}}{eBv_{\rm F}}~,
\end{equation}
where $m^*= \hbar k_{\rm F}/v_{\rm F}$ is the effective electron mass in doped graphene~\cite{geim_naturemater_2007,castroneto_rmp_2009}. 

\begin{figure}[h!]
\begin{overpic}[width=0.8\linewidth]{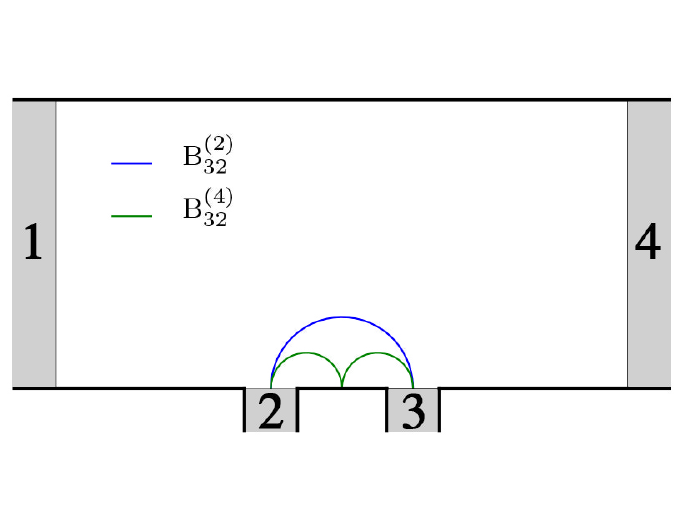}\put(2,70){(a)}\end{overpic}\\
\begin{overpic}[width=0.8\linewidth]{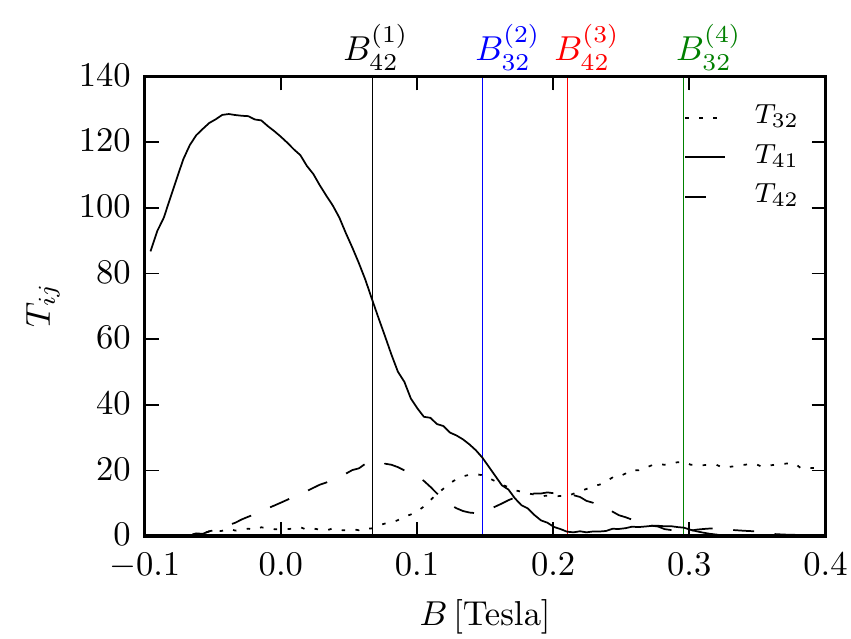}\put(2,70){(b)}\end{overpic}
\begin{overpic}[width=0.8\linewidth]{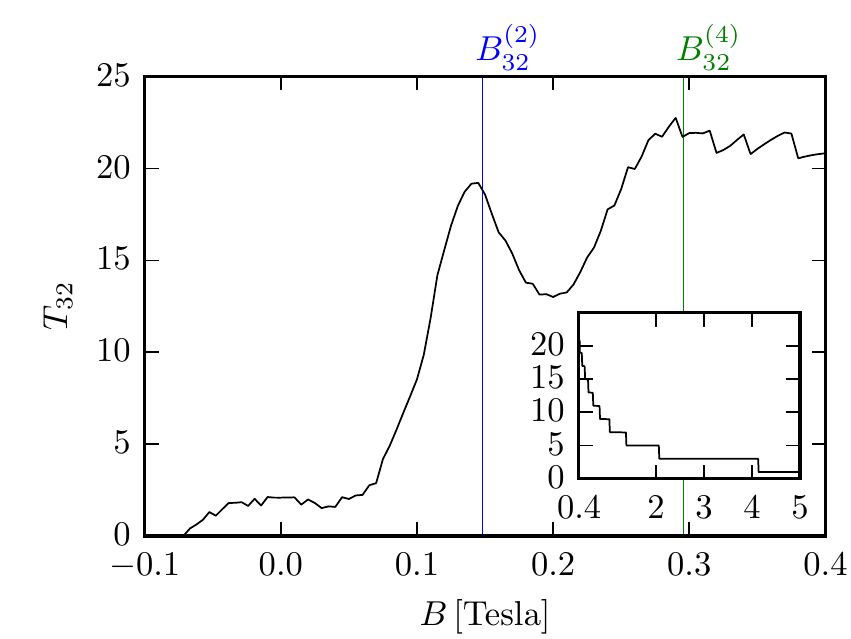}\put(2,70){(c)}\end{overpic}
\caption{\label{fig:five}
(Color online) (a) Classical electron trajectories for two values, $B_{32}^{(2)}$ and $B_{32}^{(4)}$, of the perpendicular magnetic field. (b) Numerical results for the transmission coefficients $T_{41}$ (solid line), $T_{32}$ (dotted line), and $T_{42}$ (dashed-dotted line) are plotted versus the applied magnetic field $B$ (in Tesla) for the $4$-terminal setup in Figs.~\ref{fig:one}(c) and (d), with $s=10.23$. We have plotted only three transmission coefficients since $T_{31}(B)= T_{42}(B)$. This is  because the system depicted in Fig.~\ref{fig:one}(d) is symmetric under reflection about the dotted red line in Fig.~\ref{fig:one}(d). (c) Numerical results for the transmission coefficient $T_{32}$ are plotted versus $B$ for the $4$-terminal setup in Figs.~\ref{fig:one}(c) and (d), with $s=20.05$. In the inset, $T_{32}$ is plotted for $B\geq 0.4~{\rm Tesla}$, clearly showing the transition to the integer quantum Hall regime (for the largest values of $B$ considered, $T_{32} =1$). Parameters as in Fig.~\ref{fig:three}.}
\end{figure}

We denote by $B^{(2N)}_{32}$ the field values for which the center-to-center distance $d$ between contacts 2 and 3 is an integer multiple of $2r_{\rm c}$, i.e.
\begin{equation}
	d = 2N\frac{\varepsilon_{\rm F}}{eB^{(2N)}_{32}v_{\rm F}}
\end{equation}
with $N=1,2$ corresponding to the trajectories shown in Fig.~\ref{fig:five}(a).
The transmission coefficients $T_{41}$ (solid curve), $T_{32}$ (dotted curve) and $T_{42}$ (dashed curve) are plotted, as function of $B$, in Fig.~\ref{fig:five}(b).
\begin{figure}[t]
\includegraphics[width=1.0\linewidth]{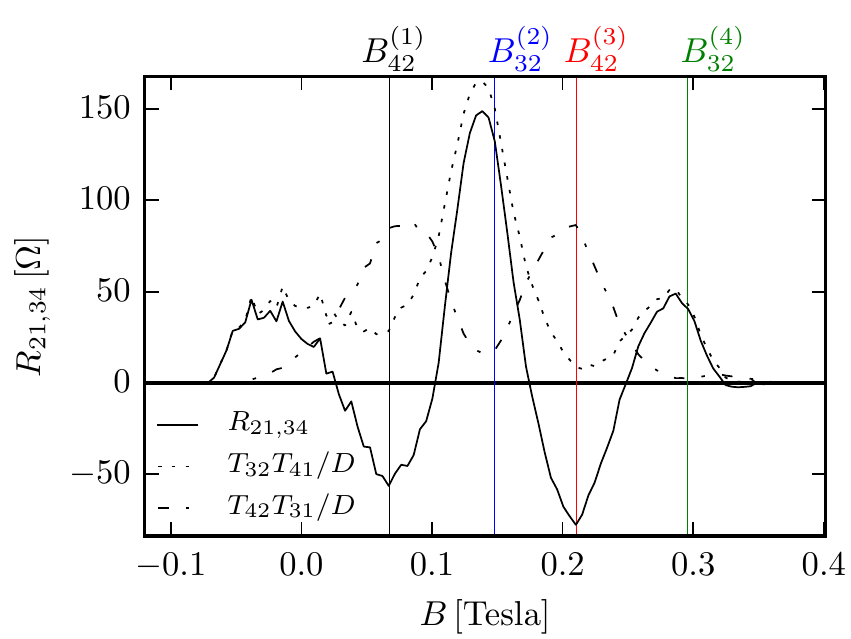}
\caption{\label{fig:six}
(Color online) Numerically calculated non-local resistance $R_{21,34}$ (solid line) versus magnetic field $B$ (in Tesla) for the $4$-terminal setup in Figs.~\ref{fig:one}(c) and (d). The dotted and dash-dotted lines represent,  respectively, the two terms $T_{32}T_{41}/D$ and $T_{42}T_{31}/D$ entering the mathematical expression of the non-local resistance $R_{21,34}$---see Eq.~(\ref{r2134}). Parameters as in Fig.~\ref{fig:three}.}
\end{figure}

Regarding the transmission probability $T_{32}$, Fig.~\ref{fig:five}(b) shows that for $B \simeq B^{(2)}_{32}$ (marked by a blue vertical line) $T_{32}$ exhibits a relative maximum, which stems from the ``direct'' trajectory with no bounces between leads $2$ and $3$ [blue line in Fig.~\ref{fig:five}(a)]. On the other hand, for magnetic fields larger than $B=B^{(4)}_{32}$ (marked by a green vertical line), $T_{32}$ exhibits a series of downward jumps---see Fig.~\ref{fig:five}(c)---preluding the eventual onset of the integer quantum Hall effect. Within the semiclassical interpretation, one expects that electrons exiting lead $2$ can either reach lead $3$ or be reflected back to lead $2$. Therefore, $T_{32}$ slowly decreases from its maximum value according to $T_{32} \simeq {\cal N}_{\rm oc} - T_{22}$, where ${\cal N}_{\rm oc}$ is the number of open channels in lead $2$. Indeed, for a fixed value of $\varepsilon_{\rm F}$, the reflection coefficient $T_{22}$ for lead $2$ increases with increasing $B$  (not shown). The slow power-law decay of $T_{32}$ for $B>B^{(4)}_{32}$ is due to trajectories with one or more bounces (skipping orbits) between leads $2$ and $3$. The trajectory with one bounce is depicted by a green line in Fig.~\ref{fig:five}(a). $T_{32}$ decreases in steps till its lowest value, $T_{32}=1$, which is reached at fields above $4~{\rm Tesla}$. In this case, leads $2$ and $3$ are connected by a single quantum Hall edge state. Note, finally, that $T_{32}$ goes to zero for large enough negative $B$, since all electrons injected from lead $2$ are in this case diverted by the Lorentz force towards lead $1$.

The transmission coefficient $T_{42}$ is characterized by a dip occurring at $B=B^{(2)}_{32}$, stemming from the fact that electrons injected from lead 2 tend to be collected mostly by lead 3 at the value of $B$ that corresponds to the cyclotron orbit connecting leads 2 and 3.
We note that $T_{42}$ goes to zero at negative values of $B$---the corresponding trajectories being deflected towards lead 1---and at large positive values of $B$---such that the small radius of the skipping orbits forces electrons injected from lead 2 to end up in the same lead. This fact together with the dip occurring at $B=B^{(2)}_{32}$ gives rise to two broad peaks in $T_{42}$ that we denote by $B^{(1)}_{42}$ and $B^{(3)}_{42}$.

Notice furthermore that $T_{31}(B)=T_{42}(B)$ since the system sketched in Fig.~\ref{fig:one}(d) is symmetric under reflection about the dotted red line in Fig.~\ref{fig:one}(d) and that $T_{41}$ is mainly characterized by a single large peak around $B=0$, since electrons injected from lead 1 have higher probability to reach lead 4 for small fields. The slight deviation of the maximum of $T_{41}$ from $B=0$ towards a negative value can be attributed to the fact that leads 2 and 3, positioned on the bottom of the Hall bar, take away electrons at small and positive values of $B$ at the expense of $T_{41}$.

As a result of Eq.~(\ref{r2134}), which expresses the non-local resistance $R_{21,34}$ in terms of the transmission probabilities, the two peaks in Fig.~\ref{fig:six} (where $R_{21,34}$ is plotted as a function of the magnetic field) at $B\simeq B^{(2)}_{32}$ and $B\simeq B^{(4)}_{32}$ stem from the two features in $T_{32}$ discussed above and are therefore genuine focusing peaks of the non-local resistance $R_{21,34}$. On the contrary, the origin of the two deep negative minima in $R_{21,34}$ is related to the two broad peaks in $T_{42}$.

We finally stress that the positivity of $R_{21,34}$ at $B\approx 0$ is due to the large value---see Fig.~\ref{fig:five}(b)---of $T_{41}$ for small (positive and negative) values of $B$, which originates from the fact that leads $1$ and $4$ are much wider than leads $2$ and $3$. Note, however, that an additional contact---such as terminal $5$ in Fig.~\ref{fig:one}(a) and~(b)---present on the upper side of the Hall bar can serve as an electron drain. This may significantly affect the discussed picture at $B\approx 0$ assuming that $W$ is much smaller than the mean free path $\ell$, so that even negative values of $R_{21,34}$ can be found, depending on the relative size and position of the extra contact. However, if $W$ is larger than $\ell$, negative values of the non-local resistance $R_{21,34}$ (termed ``vicinity'' resistance in Ref.~\onlinecite{bandurin_science_2016}) in zero magnetic field cannot be explained within a single-particle ballistic approach~\cite{bandurin_science_2016}. As we will see below in Sect.~\ref{sect:disorder}, elastic disorder at the edges is not able to change the clean-limit picture. Negative values of $R_{21,34}$ (which occur only at sufficiently large temperatures) have been attributed to hydrodynamic viscous flow~\cite{bandurin_science_2016,torre_prb_2015}.

To further emphasize the relation between transmission coefficients and non-local resistance, in Fig.~\ref{fig:six} we plot separately the two terms appearing in Eq.~(\ref{r2134}) along with $R_{21,34}$.
The plot makes clear that the term $T_{32}T_{41}/D$ determines the occurrence of the positive peaks, while the term $T_{42}T_{31}/D$ is responsible for the appearance of the two negative dips. This interpretation of the negative dips is in agreement with earlier theoretical work~\cite{milovanovic_jap_2014}, based on a semiclassical billiard model, where TMF in a geometry identical to that in Figs.~\ref{fig:one}(c) and (d) was discussed.

\begin{figure}[t]
\begin{overpic}[width=\linewidth]{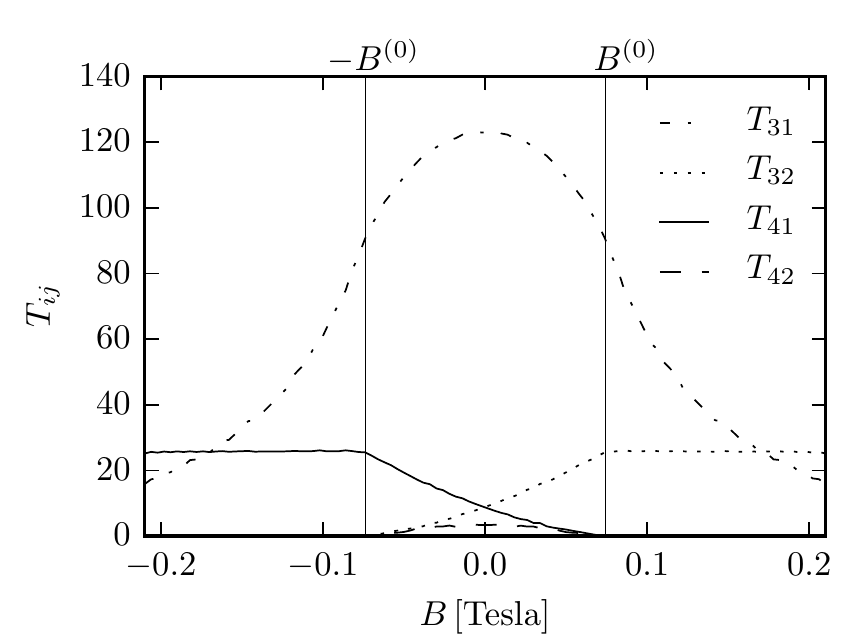}\put(2,70){(a)}\end{overpic}\\
\begin{overpic}[width=\linewidth]{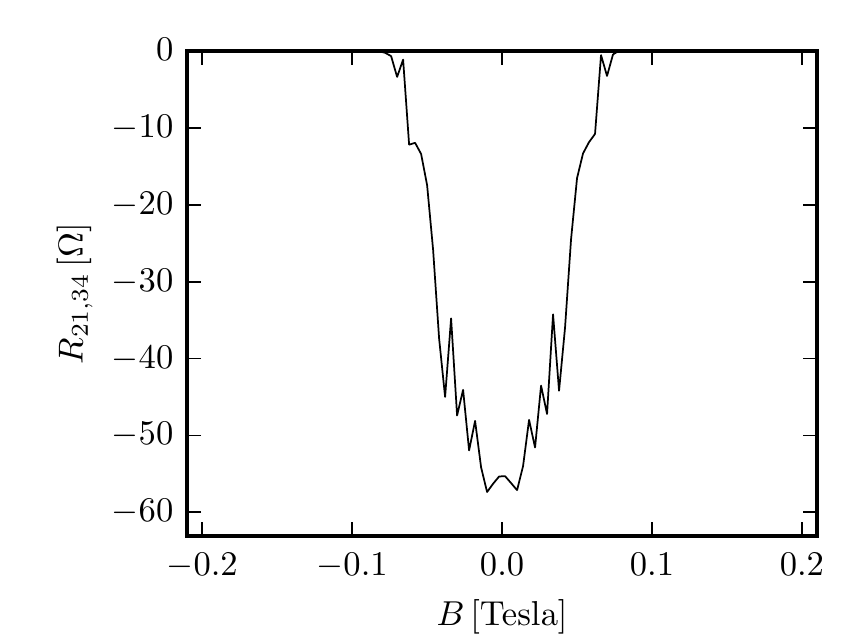}\put(2,70){(b)}\end{overpic}
\caption{\label{fig:seven}
(a) Numerically calculated transmission coefficients $T_{41}$ (solid line), $T_{32}$ (dotted line), $T_{42}$ (dashed line), and $T_{31}$ (dashed-dotted line) are plotted versus the applied magnetic field $B$ (in Tesla) for the $4$-terminal setup in Fig.~\ref{fig:one}(e) and (f), with $s=10.23$. Note that $T_{41}(-B)=T_{32}(B)$  because the system depicted in Fig.~\ref{fig:one}(f) has an inversion symmetry center, marked by a red cross in Fig.~\ref{fig:one}(f).
(b) Non-local resistance $R_{21,34}$ relative to the transmissions in panel (a). Parameters as in Fig.~\ref{fig:three}.}
\end{figure}

We now turn to an analysis of non-local ballistic magneto-transport in the 4-terminal setup sketched in 
Fig.~\ref{fig:one}(e) and (f), with the two vertical leads placed on {\it opposite} sides of the Hall bar. 
The relevant transmission coefficients are plotted in Fig.~\ref{fig:seven}(a) as functions of the magnetic field.
Note that $T_{31}$ (dash-dotted line) and $T_{42}$ (dashed line) respect the following symmetry: $T_{31}(B) = T_{31}(-B)$ and $T_{42}(B)= T_{42}(-B)$. Also,  we note that $T_{32}(B)= T_{41}(-B)$. This is because the system depicted in Fig.~\ref{fig:one}(f) has an inversion symmetry center, marked by a red cross in Fig.~\ref{fig:one}(f). The transmission coefficient between the two widest electrodes of the system, $T_{31}$, is however much larger than $T_{42}$ and shows a smooth bell-like shape, which decreases slowly with increasing magnitude of $B$.
On the contrary, $T_{42}$ shows spiky features (not visible on the scale of the plot), possibly arising from quantum interference effects, and goes rapidly to zero at a value of the field ($|B|\simeq B^{(0)}$) that yields a cyclotron radius equal to $W/2$, i.e.
\begin{equation}
B^{(0)}= \frac{2\varepsilon_{\rm F}}{e Wv_{\rm F}}~.
\end{equation}
This is due to the fact that electrons injected from lead 2 cannot reach lead 4 for $B\simeq \pm B^{(0)}$, being deflected towards lead 3 (lead 1).
This is confirmed by the behavior of $T_{32}$ (dotted line), which increases for increasing $B$, reaching a constant value for $B\geq B^{(0)}$, close to the number of open channels in lead 2 (${\cal N}_{\rm oc}=27$).
Notice also that $T_{32}$ is negligible only when $B\leq -B^{(0)}$.
The non-local resistance $R_{21,34}$ can be calculated using Eq.~(\ref{r2134}). Numerical results are reported in Fig.~\ref{fig:seven}(b) as a function of $B$. It turns out that $R_{21,34}$ resembles very closely the shape of $T_{42}$, but with a negative sign since $R_{21,34}$ is dominated by the term $T_{42}T_{31}/D$ at all values of $B$.
It is worthwhile noticing that in our simulations $R_{21,34}$ is never positive since for $B\leq -B^{(0)}$, when $T_{42}$ becomes negligible, $T_{32}$ gets negligible too.

\subsection{Classical trajectory model}
\label{sect:classical}
To further investigate the classical nature of the main features in the non-local resistance $R_{21,34}$, we have developed a model based on fully classical trajectories, which allows us to calculate the transmission probabilities between electrodes.

\begin{figure}[t]
\includegraphics[width=1.0\linewidth]{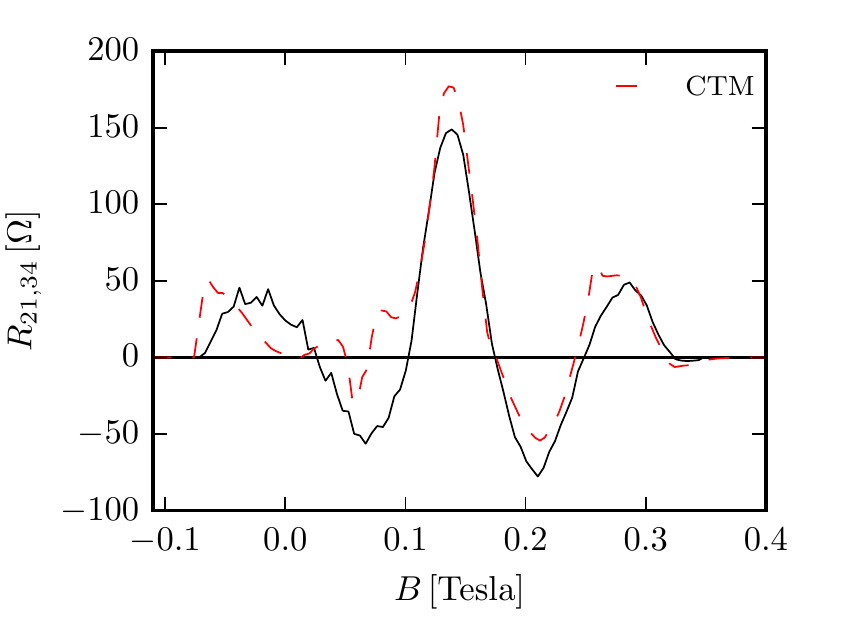}
\caption{\label{fig:eight}
(Color online) Numerically calculated non-local resistance $R_{21,34}$ for the $4$-terminal setup in Fig.~\ref{fig:one}(c) and (d) as a function of $B$ calculated using the classical trajectory model (CTM, red dashed line). The black line is the result obtained with the quantum tight-binding model (same curve plotted in Fig.~\ref{fig:six}).
We take the following parameters: $M_{\rm e}=100$ and $M_{\rm p}=500$.
Sample parameters are the same as in Fig.~\ref{fig:three}.}
\end{figure}

The model is detailed as follows:
\begin{itemize}
\item We assume that electrons move in the Hall bar according to the classical equations of a charged particle in a transverse magnetic field;
\item In a given electrode, electrons are emitted from $M_{\text{p}}$ equidistant points;
\item From each such point, $M_{\text{e}}$ electrons are emitted with an {\it isotropic} distribution of angles with a fixed magnitude of velocity (equal to the Fermi velocity $v_{\rm F}$);
\item The number of electrons $M_{ij}$ (with $i,j=1,2,3,4$) arriving in electrode $i$ when emitted from electrode $j$ is determined by the classical equations (notice that $M_{ij}\leq M_{\text{p}} M_{\text{e}}$);
\item The transmission probabilities $\overline{T}_{ij}$ are defined by normalizing the coefficients $M_{ij}$ as follows:
\begin{equation}
\overline{T}_{ij} = \alpha_i \beta_j M_{ij}~,
\end{equation}
where $\alpha_i$ and $\beta_j$ are numerical coefficients determined by imposing the following conditions
\begin{equation}
\sum_{i} \overline{T}_{ij}= N_j
\label{c1}
\end{equation}
and
\begin{equation}
\sum_{j} \overline{T}_{ij}= N_i~.
\label{c2}
\end{equation}
Here, $N_i$ is the number of open channel in lead $i$ as defined in the tight-binding quantum model, see Sec.~\ref{sect:TB};
\item The non-local resistance $R_{21,34}$ is finally calculated substituting the transmission probabilities $\overline{T}_{ij}$ in Eq.~(\ref{r2134}).
\end{itemize}
Notice that the conditions (\ref{c1}) and (\ref{c2}) express particle current conservation within the scattering approach used in the quantum model of Sect.~\ref{sect:TB}.
In Fig.~\ref{fig:eight} we plot the non-local resistance obtained with this method as a function of $B$ (red dashed line), along with the result obtained with the quantum model of Sect.~\ref{sect:TB} (black solid line).
Fig.~\ref{fig:eight} shows that the main features of $R_{21,34}$, in particular the two peaks for $B	>0$ and the two negative minima, are well reproduced by the classical trajectory model.
This result confirms the classical nature of the main features of $R_{21,34}$, phase coherence playing a little role.

\subsection{Carrier density dependence}
\label{sect:density}
So far we have seen that the main features of the non-local resistance can be explained on a classical level.
This is due to the fact that the value of the Fermi energy used for the plot in Fig.~\ref{fig:six}, $\varepsilon_{\rm F} = 66.86~{\rm meV}$, corresponds to the relatively highly doped graphene sheet used in the measurements (see below).
One expects, however, that quantum effects become more important by decreasing the carrier density (i.e.~the Fermi energy), thus moving to a regime where a few electronic modes are involved in transport. Upon decreasing density, however, also disorder becomes important, which we here have decided to neglect.

\begin{figure}[t]
\begin{overpic}[width=0.49\linewidth]{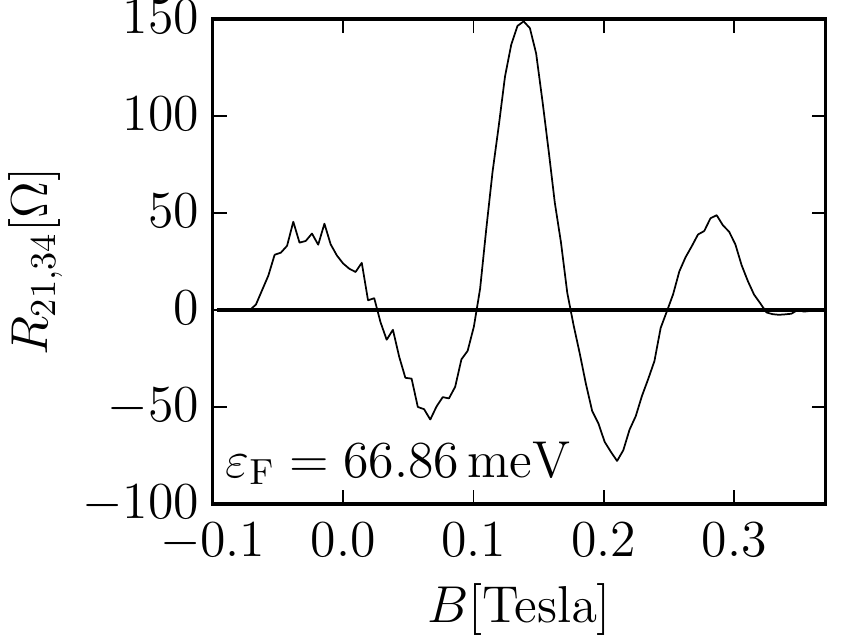}\put(0,72){(a)}\end{overpic}
\begin{overpic}[width=0.49\linewidth]{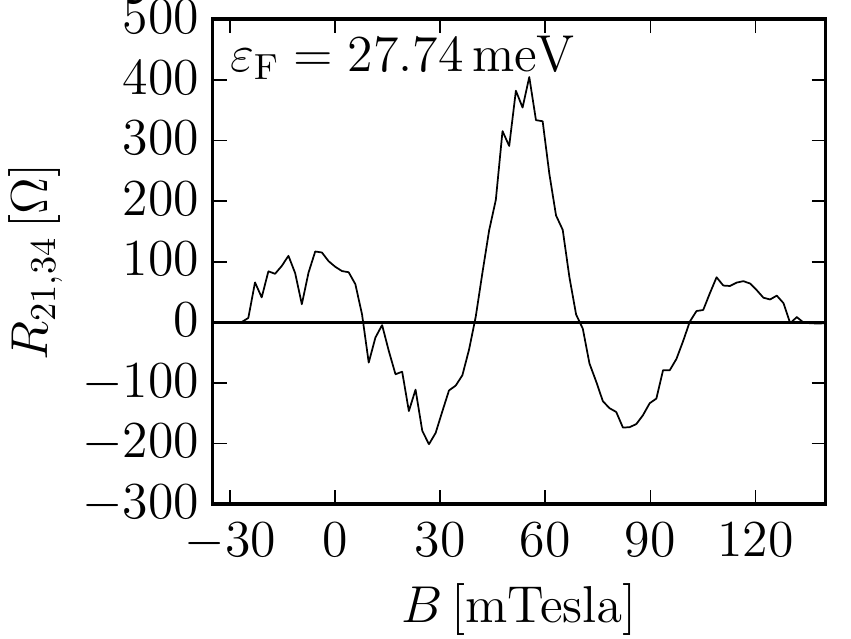}\put(0,72){(b)}\end{overpic}
\begin{overpic}[width=0.49\linewidth]{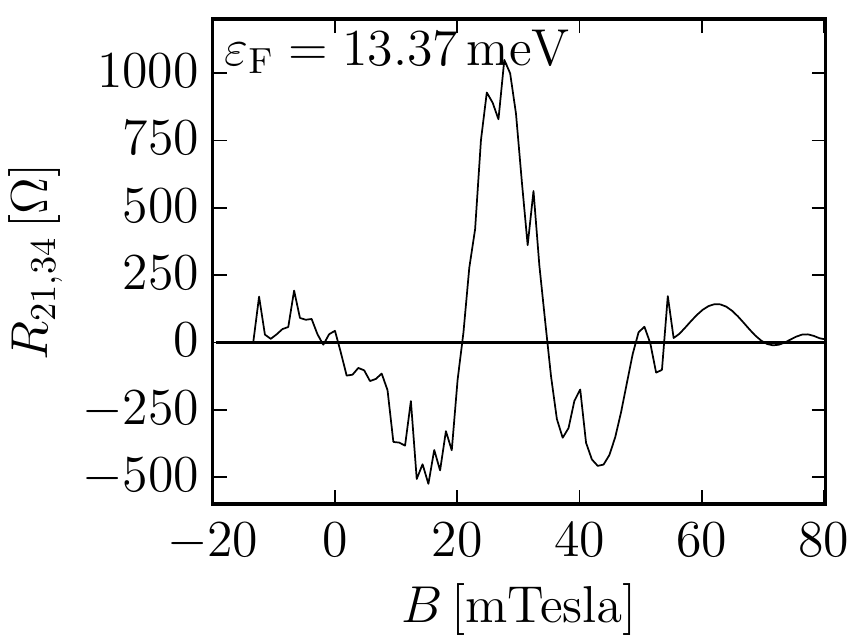}\put(0,72){(c)}\end{overpic}
\begin{overpic}[width=0.49\linewidth]{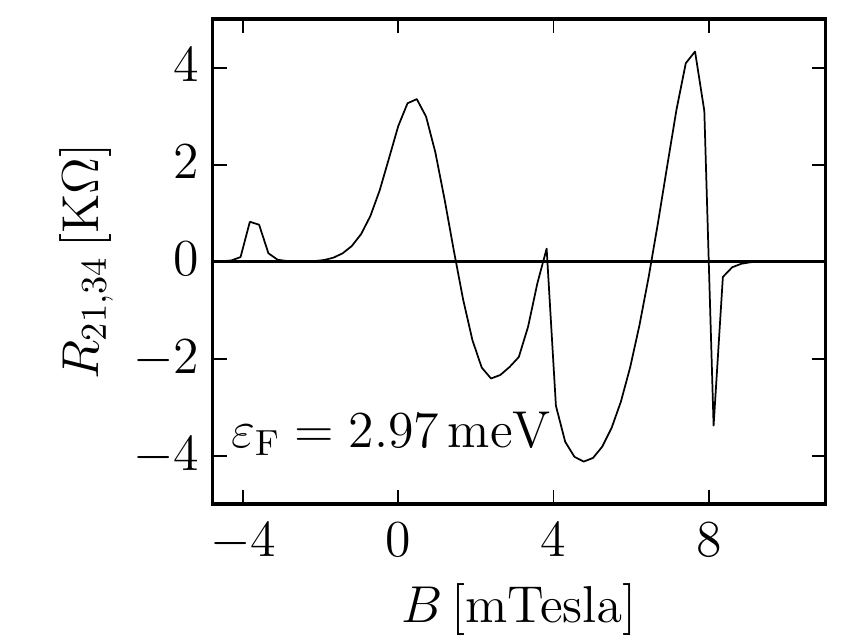}\put(0,72){(d)}\end{overpic}
\caption{\label{fig:nine} Numerically calculated non-local resistance $R_{21,34}$ versus applied magnetic field $B$ at different values of the Fermi energy [(a) $\varepsilon_{\rm F}= 66.86$ meV, (b) $\varepsilon_{\rm F}= 27.74$ meV, (c) $\varepsilon_{\rm F}= 13.37$ meV, and (d) $\varepsilon_{\rm F}= 2.97$ meV] for the $4$-terminal setup in Fig.~\ref{fig:one}(c) and (d). The scaling factor used for the calculations is $s=10.23$, while the sample parameters are the same as in Fig.~\ref{fig:three}.
The number of open channels in the leads depend on $\varepsilon_{\rm F}$. In leads $2$ and $3$ the number of open channels is: (a) ${\cal N}_{\rm oc}=27$, (b) ${\cal N}_{\rm oc}=11$, (c) ${\cal N}_{\rm oc}=5$, and (d) ${\cal N}_{\rm oc}=1$.
Notice that the scales (in both the resistance and field axes) are different in the various panels.
The focusing peaks remain located where predicted by the classical analysis in panel (a), (b) and (c).}
\end{figure}

Fig.~\ref{fig:nine} shows the evolution of the non-local resistance versus magnetic field, at zero temperature, as the Fermi energy is decreased.
Starting from Fig.~\ref{fig:nine}(a), relative to $\varepsilon_{\rm F} = 66.86~{\rm meV}$, one observes that the value of the resistance increases while the focusing peaks ``degrade'', but still persist for $\varepsilon_{\rm F} = 27.74~{\rm meV}$ [Fig.~\ref{fig:nine}(b)] and for $\varepsilon_{\rm F} = 13.37~{\rm meV}$ [Fig.~\ref{fig:nine}(c)], with peak positions shifting in agreement with Eq.~(\ref{eq:cyclotronradius}).
By further lowering $\varepsilon_{\rm F}$, Fig.~\ref{fig:nine}(d) shows that for $\varepsilon_{\rm F} = 2.97~{\rm meV}$ the non-local resistance presents a completely different structure which cannot be understood in classical terms.
Notice, in particular, that in this latter case the number of open channels in leads 2 and 3 ($N_2$ and $N_3$), the narrowest in the system, is equal to $1$.
Since  focusing peaks are still distinguishable when $N_2=N_3=5$---$\varepsilon_{\rm F} = 13.37~{\rm meV}$ 
as in Fig.~\ref{fig:nine}(c)---we can conclude that the quantum regime sets in when the number of open channels is close to $1$. A complete analysis of the quantum regime and the interplay between electron-hole puddles and quantum interference, though, is beyond the scope of the present Article.

\subsection{Thermal smearing of the Fermi surface}
\label{sect:temperature}

In this Section we analize the impact of the smearing of the Fermi surface due to finite-temperature effects on TMF.

Within the scattering approach in the linear-response regime, the effect of a finite temperature $T$ is taken into account by replacing in Eqs.~(\ref{r2134}-\ref{last}) the transmissions probabilities $T_{ij}$, evaluated at the Fermi energy, with the following energy integrals
\begin{equation}
\langle T\rangle_{ij}=\int_{-\infty}^{\infty}T_{ij}(E)\left( -\frac{\partial f(E)}{\partial E} \right)dE~,
\label{finitet}
\end{equation}
where $f(E)=[\exp(E/(k_{\rm B} T))+1]^{-1}$ is the Fermi distribution function at temperature $T$. 

\begin{figure}[t]
\includegraphics[width=1.0\linewidth]{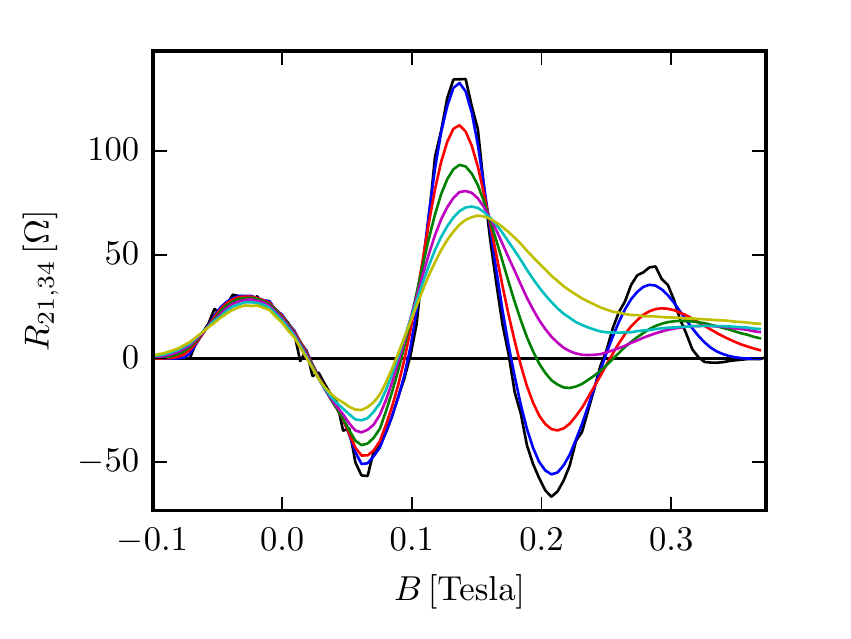}
\caption{\label{fig:ten}
(Color online) Numerically calculated non-local resistance $R_{21,34}$ versus applied magnetic field $B$ at different temperatures $T = 0,\, 25,\, 50,\, 75,\, 100,\, 125,\, 150~{\rm K}$ for the system depicted in Fig.~\ref{fig:one}(c) and (d). The scaling factor used for the simulations is $s=20.05$ and all other parameters are the same as for Fig.~\ref{fig:three}. Note that the temperature needs to be rescaled with the energy scaling factor $s'$. Data in this plot are just meant to give the reader an idea of the magnitude of Fermi-surface smearing effects on TMF signals. As shown in Ref.~\onlinecite{bandurin_science_2016}, non-local electrical signals at temperatures $T\gtrsim T_{\rm hydro}$, defined in Sect.~\ref{sect:intro}, are sensitive to electron-electron interactions, which are not included in the numerical calculations presented in this work.}
\end{figure}

Plots of the non-local resistance $R_{21,34}$ as a function of the magnetic field $B$ and for different values of $T$ are reported in Fig.~\ref{fig:ten}.
As expected, the non-local resistance becomes smoother for increasing values of $T$ and the height of the focusing peaks decreases as $T$ increases. Notice, however, that the first peak for positive values of $B$, which is not related to focusing, is hardly affected by temperature. This behavior can be understood on classical terms from the fact that, at finite temperatures, electrons contributing to $\langle T\rangle_{ij}$ are emitted at different energies, according to Eq.~(\ref{finitet}), and thus move at different cyclotron radii.
More precisely, the values of the cyclotron radii will be distributed around the zero-temperature value [Eq.~(\ref{eq:cyclotronradius})] with a width proportional to  temperature and given by
\begin{equation}
\delta r_c = \frac{k_{\rm B}T}{eBv_{\rm F}}~.
\end{equation}
In other words, with increasing temperature a larger range of values for the cyclotron radius contributes to all transmissions $\langle T\rangle_{ij}$ so that they get non-vanishing on a larger interval of values of $B$.
As a result, the focusing effect is blurred. The non-local resistance diminishes in magnitude at all fields with increasing temperature and remains finite for larger values of $B$.

Note that in Fig.~\ref{fig:ten} the peaks occurring at $B=B^{(1)}_{42}$ and $B=B^{(2)}_{32}$ remain distinguishable at all temperatures, although the decrease of their height is nearly exponential with $T$.
The peaks occurring at $B=B^{(3)}_{42}$ and $B=B^{(4)}_{32}$, however, are more strongly affected, eventually disappearing for the largest temperatures considered.

In this Section we have analyzed only Fermi-surface smearing effects induced by a finite temperature. In reality, also inelastic collisions between electrons and agents external to the 2D electron system (e.g.~acoustic phonons) play a role in determining the magnitude of the non-local signal in a TMF experiment. Our results in Fig.~\ref{fig:ten} clearly show that Fermi-surface smearing effects play a non-negligible role and must be taken into account in any serious comparison between microscopic theoretical predictions and experiments.

\subsection{Experimental data versus numerical calculations}
\label{sect:exp}

We have carried out transport experiments on Hall bar devices with two current and four potential probes (two potential probes on each side). 
To achieve mean-free paths $\ell$ larger than the sample size, graphene was encapsulated in hexagonal boron nitride~\cite{mayorov_nanolett_2011}. Fabrication details can be found in the Supplementary Material of Ref.~\onlinecite{bandurin_science_2016}.

The characteristic geometrical details of our devices (Hall bar width $W$, distance $d$ between current and potential probes, and width $w$ of the probes) are the same 
as in the numerical calculations discussed in Sect.~\ref{sect:numerical_examples}. A standard low-frequency AC technique  was employed for measurements of the $B$-field dependence of the 4-probe resistance in a commercial cryostat with a superconducting magnet.

\begin{figure}[t]
\begin{overpic}[width=1.0\linewidth]{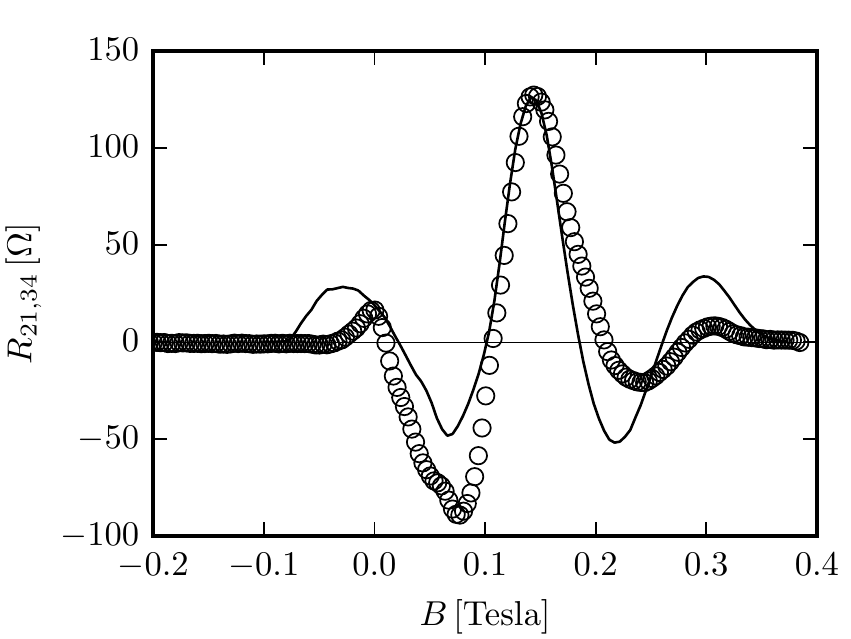}\put(0,72){(a)}\end{overpic}
\begin{overpic}[width=1.0\linewidth]{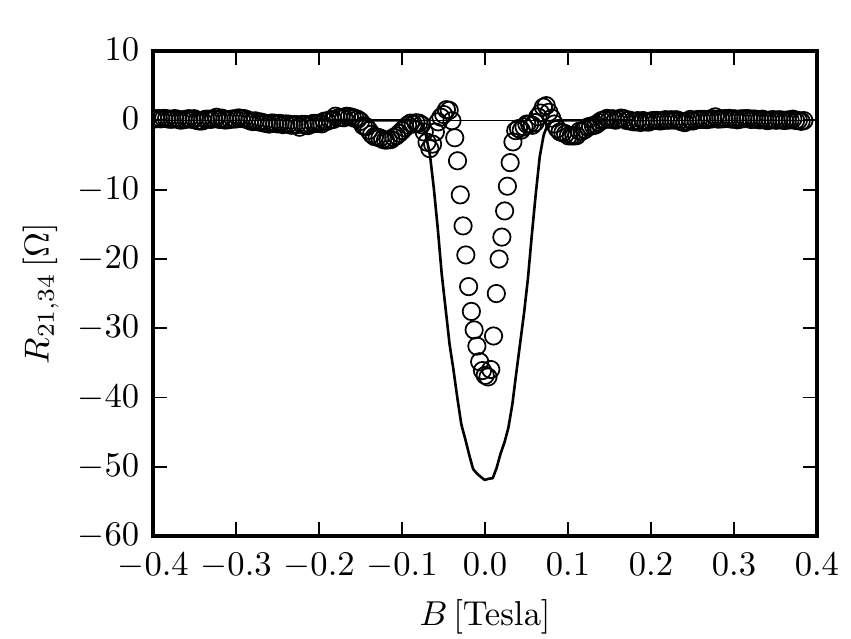}\put(0,72){(b)}\end{overpic}
\caption{Comparison between experimental results (empty circles) 
and results from numerical calculations (solid line)  for the non-local resistance $R_{21,34}$ at $T = 25~{\rm K}$. 
Panel (a) refers to the setup in Fig.~\ref{fig:one}(c) while panel (b) refers to the setup in Fig.~\ref{fig:one}(e).
Sample parameters are $W=2$ $\mu$m, $L_1=1.5$ $\mu$m, $L_2=3.5$ $\mu$m, $w=0.37$ $\mu$m, $d=1~{\rm \mu m}$, and a carrier density $n = 0.4\times 10^{12}~{\rm cm}^{-2}$.\label{fig:eleven} }
\end{figure}

Typical TMF experimental traces are shown in Fig.~\ref{fig:eleven}. It compares the measurements with our Landauer-B\"{u}ttiker calculations. In order to do so, we use the parameters reported in the caption of Fig.~\ref{fig:eleven}.
As for the scaling factor, we use $s=20.05$, while the value of the rescaled Fermi energy has been slightly adjusted, with respect to the value $\tilde{\varepsilon}_{\rm F}$ dictated by the scaling procedure, in order to fit the position of the main peak in the non-local resistance. The value used for the numerical calculations is $\tilde{\varepsilon}'_{\rm F} = 1.37~{\rm eV}$, whereas the value obtained from the scaling procedure is $\tilde{\varepsilon}_{\rm F}=1.31~{\rm eV}$, thus differing only by less than $5\%$. This adjustment is justified by the fact that the value of the Fermi energy $\varepsilon_{\rm F}$---see input box in Fig.~\ref{fig:two}---is inferred from the experimental value of the carrier density $n$, assuming the usual massless Dirac fermion relation $\varepsilon_{\rm F}= \hbar v_{\rm F}\sqrt{\pi |n|}$, which is only approximately valid for the vertical lead of width $w=0.37~{\rm \mu m}$ that is used in our algorithm to calculate the number of open channels ${\cal N}_{\rm oc}$. Such small discrepancies, $\lesssim 5~\%$, may stem from a variety of reasons including the nature of edges (zig-zag, armchair, or a combination), electron-hole aysmmetry~\cite{kretinin_prb_2009}, quantum confinement, etc. Note, moreover, that we allow only $\tilde{\varepsilon}_{\rm F}$ as a ``fit'' parameter, while taking as the hopping energy value (see discussion in Sect.~\ref{sect:TB}) its bare non-interacting tight-binding value $t_{0}$.

In Fig.~\ref{fig:eleven} the measured non-local resistance $R_{21,34}$ (empty circles) as a function of $B$ is plotted along with the numerical result (solid line) for the 4-terminal setups in Fig.~\ref{fig:one}(c) and (d), panel (a), and in Fig.~\ref{fig:one}(e) and (f), panel (b). The comparison reveals good agreement such that the main features of $R_{21,34}$ are reproduced as well as its absolute value.
In particular, the main peak in Fig.~\ref{fig:eleven}(a) is nearly perfectly reproduced, while the right peak is in the correct position, although exhibiting a smaller height. The position and shape of the left dip is also well captured, but not its amplitude.
Regarding Fig.~\ref{fig:eleven}(b), our calculations reproduce the presence of a single minimum at zero field, but with a larger amplitude and with no additional oscillations.
These discrepancies may be imputed to the actual detailed structure of the sample, disorder, and other non-idealities.

\subsection{The role of non-ideal edges}
\label{sect:disorder}

In this Section we discuss the consequences of possible imperfections present at the edges of the Hall bar. We focus on their impact on the non-local resistance $R_{21,34}$ at $B=0$. Our aim here is to show that the conclusions drawn above in Sect.~\ref{analy} on the positivity of $R_{21,34}$ at $B=0$ are robust against structural disorder at the edges.

\begin{figure}[t]
\begin{overpic}[width=0.8\linewidth]{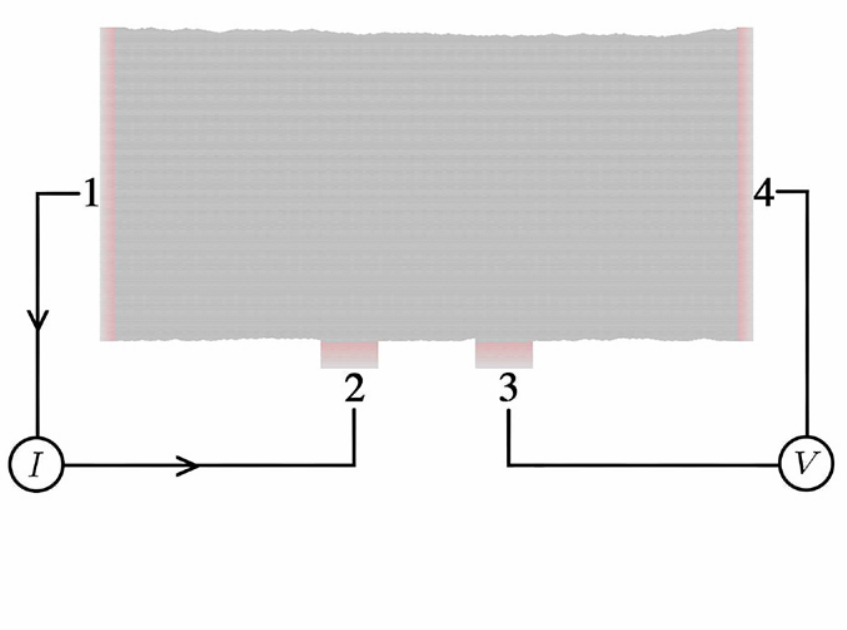}\put(0,72){(a)}\end{overpic}
\begin{overpic}[width=0.8\linewidth]{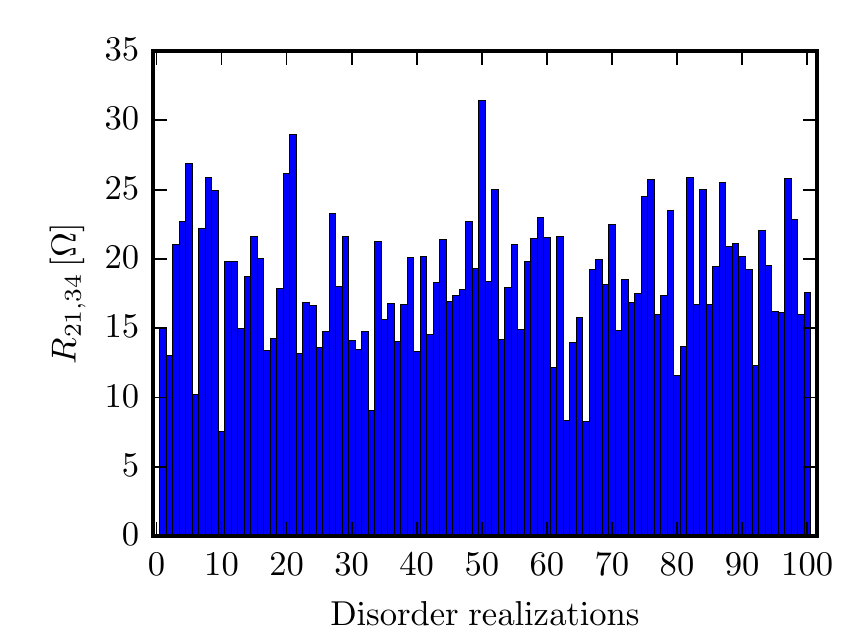}\put(0,72){(b)}\end{overpic}
\begin{overpic}[width=0.8\linewidth]{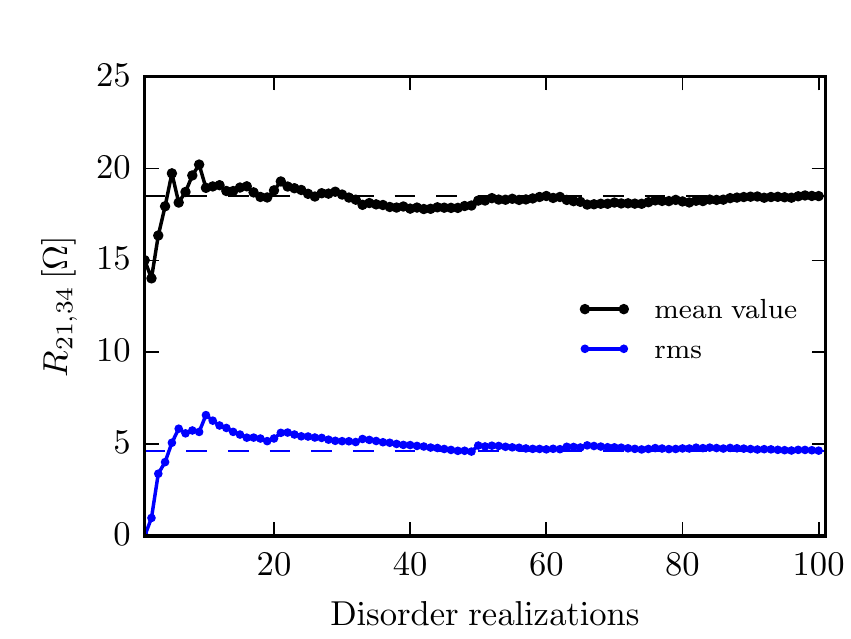}\put(0,72){(c)}\end{overpic}
\caption{(Color online) Numerical results for the non-local resistance in zero magnetic field and for non-ideal edges. (a) Example of a graphene ribbon with non-ideal edges, with $M_{\text{R,min}}=2$ and $M_{\text{R,max}}=6$. (b) Histogram of the non-local resistances at zero magnetic field ($B=0$) 
obtained for $100$ different random configurations. The relative mean value is $\overline{R}_{21,34}=18.50~\Omega$ with standard deviation equal to $\Delta R_{21,34}=4.65~\Omega$. (c) Mean value and standard deviation as a function of the number of random configurations.
The scaling factor used for the simulations is $s=20.05$, $M_{\text{R,min}}=2$ and $M_{\text{R,max}}=6$, and all other parameters are the same as in Fig.~\ref{fig:three}. \label{fig:twelve} }
\end{figure}

Edge imperfections are implemented by carving independently the two horizontal edges using an algorithm which, at random, adds or removes two rows of atoms from each sublattice (taking care of avoiding dangling bonds) of length corresponding to a number of sites $M_{\text{R}}$, which is also randomly chosen in the range $[M_{\text{R,min}},M_{\text{R,max}}]$.
An example of the resulting nanoribbon is presented in Fig.~\ref{fig:twelve}(a).
A constraint is imposed on the maximum nanoribbon width, which is set by $W$.
The histogram in Fig.~\ref{fig:twelve}(b) shows the values obtained for the non-local resistance of 100 different random configurations for $M_{\text{R,min}}=2$ and $M_{\text{R,max}}=6$.
The mean value turns out to be $\overline{R}_{21,34}=18.50$ $\Omega$ with standard deviation equal to $\Delta R_{21,34}=4.65$ $\Omega$ (for comparison, recall that for the corresponding ideal nanoribbon one finds $R_{21,34}=20.69$ $\Omega$, well within a standard deviation).
Fig.~\ref{fig:twelve}(c), on the other hand, shows the evolution of the mean value and standard deviation of non-local resistance with increasing number of random configurations, proving that convergence is obtained already with 60 configurations.
We additionally mention that mean value and standard deviation of $R_{21,34}$ do not significantly change if $M_{\text{R}}$ varies in a larger range of values.
Namely, for $M_{\text{R,min}}=4$ and $M_{\text{R,max}}=10$ we find $\overline{R}_{21,34}=19.72$ $\Omega$ and $\Delta R_{21,34}=6.49$ $\Omega$.

\section{Conclusions}
\label{sect:conc}
In this report we have proposed a scaling procedure, based on the tight-binding approach and Landauer-B\"{u}ttiker theory, for transport calculations in ultra-clean graphene devices of realistic size.

The procedure is based on the exact band structure of graphene {\it ribbons}, and uses the Fermi energy as key scaling parameter.
We have demonstrated the effectiveness of the procedure by calculating the non-local resistance of a realistic $5$-terminal setup in the presence of a magnetic field.
In such a transverse magnetic focusing setup, we have compared the non-local resistance as a function of magnetic field for increasing values of the scaling factor, proving that this approach is particularly suitable for micron-sized ribbons and in the presence of many electrodes.

The case of transverse magnetic focusing has been further analysed in realistic $4$-terminal setups, where the structure of the non-local resistance as a function of magnetic field has been explained in terms of classical cyclotron orbits.
Moreover, we have addressed the dependence of the non-local resistance on the carrier density and temperature and studied the impact of disorder at the edges of the ribbon.

Finally, we have compared the results of our scaling approach with experimental data in high-quality encapsulated samples finding good agreement.
The main features, as well as the absolute value of the non-local resistance, are well reproduced using the actual experimental parameters.

\acknowledgements 
This work was supported by the EU Horizon 2020 research and innovation programme under grant agreement no.~696656 ``GrapheneCore1'', the EU project ``ThermiQ", the EU project COST Action MP1209 ``Thermodynamics in the quantum regime", the EU project COST Action MP1201 ``NanoSC", and the SNS internal project ``Thermoelectricity in nanodevices".

Free software (www.gnu.org, www.python.org) was used.

\end{document}